\documentclass[aps,prl,nofootinbib,reprint,superscriptaddress]{revtex4-1}
\usepackage{blindtext}
\usepackage{amsmath}
\usepackage{amsfonts}
\usepackage{amssymb}
\usepackage{hyperref}

\begin{document}

\title{\bf Generalized entanglement temperature and entanglement Smarr relation}
\vskip 1cm
\author{Ashis Saha}
\email{sahaashis0007@gmail.com}
\affiliation{Department of Physics, University of Kalyani, Kalyani 741235, India}
\author{Sunandan Gangopadhyay}
\email{sunandan.gangopadhyay@bose.res.in}
\affiliation{Department of Theoretical Sciences,
S.N.~Bose National Centre for Basic Sciences, JD Block, Sector-III, Salt Lake, Kolkata 700106, India}
\author{Jyoti Prasad Saha}
\affiliation{Department of Physics, University of Kalyani, Kalyani 741235, India}



\begin{abstract}
\noindent We observe that in presence of excitation, a thermodynamic Smarr like relation corresponding to a generalized entanglement temperature ($T_g$) can be holographically obtained for the entanglement entropy of a subsystem. Such a relation emerges naturally by demanding that the generalized entanglement temperature produces the exact Hawking temperature as the leading term in the IR limit ($l\rightarrow \infty$). Remarkably, this relation has the same form as the Smarr relation in black hole thermodynamics. We demonstrate this for three spacetime geometries, namely, a background with a nonconformal factor, a hyperscaling violating geometry background, and a charged black hole background which corresponds to a field theory with a finite chemical potential.
\end{abstract}
\maketitle



\noindent The entropy of a thermodynamical system counts the number of microstates (quantum mechanical information) of the system and is related to internal energy of the concerned system via the first law of thermodynamics $dE=T~dS$. On the other hand,  entanglement entropy is a good measurement of quantum entanglement for a pure state \cite{nielsen,Calabrese}. It has been a matter of great interest to look for a thermodynamic relation for the entanglement entropy of a system when it is under excitation. Such a relation was first obtained in the context of gauge/gravity correspondence \cite{maldacena}, where it was observed that the holographic entanglement entropy (HEE) of a system in the ultra-violet (UV) limit satisfies a thermodynamic like relation \cite{jb}. Further, this observation led to the concept of entanglement temperature $T_{ent}$, which has the universal behaviour of being inversely proportional to the size of the concerned subsystem in the field theory. 
A natural question is whether such a thermodynamics like relation for the HEE holds for all possible values of the subsystem size $l$.
It is also important to note that the thermodynamics like law obeyed by the HEE holds only in the form of a Smarr relation.
Interestingly, in \cite{prd} a thermodynamics like law was obtained for the HEE in arbitrary dimensions. A generalized entanglement temperature $T_g$ was introduced and it was required to give the Hawking temperature in the IR limit. This also produced the entanglement temperature in the UV limit. This makes it natural to look for a thermodynamics like law for HEE in various other scenarios, namely, a more general class of field theories which do not enjoy conformal symmetry, hyperscaling violating theories and theories with a finite chemical potential. As mentioned earlier, the principal drawback of the concept of $T_{ent}$ (introduced in \cite{jb}) lies in the fact that it is valid only in the UV limit ($l\rightarrow 0$) in the holographic set up and is not able to probe the behaviour in the IR limit ($l\rightarrow \infty$). In this letter we resolve this drawback by introducing a generalized entanglement temperature $T_g$, valid for the whole subsystem length. We further impose a physical requirement that $T_g$ reduces to the Hawking temperature of the black hole in the IR limit. The justification for this requirement is the following. In the holographic scenario, the static minimal surface corresponding to a thermally excited quantum field theory, wraps a portion of the event horizon, capturing flux of the Hawking radiation in the IR limit, and therefore $T_g$ on the static minimal surface should reduce to the Hawking temperature of the black hole in the IR limit. Remarkably, this physical requirement immediately leads to a thermodynamics like Smarr relation obeyed by the HEE valid for the whole subsystem length. 
The expression for the generalized temperature is valid along the whole scale for the dual field theory and reproduces the universal behaviour of the entanglement temperature in the UV limit with the appropriate numerical factor.\\
We start our discussion by considering a Schwarzschild-type black brane solution in the Einstein-dilaton gravity theory with Liouville potential in ($d+1$) spacetime dimensions. The importance of this background lies in the fact that the dual field theory of this gravity model is relativistic and nonconformal in nature. The metric of this gravity theory reads \cite{cpark1}
\begin{eqnarray}\label{1}
ds^2&=&-r^{2p}f(r)dt^2+\frac{dr^2}{r^{2p}f(r)}+r^{2p}\sum_{i=1}^{d-1}dx_i^2;\\
f(r)&=&1-\left(\frac{r_h}{r}\right)^k,p=\frac{8}{8+(d-1)\eta^2}, k= \frac{8d-(d-1)\eta^2}{8+(d-1)\eta^2}\nonumber.
\end{eqnarray}
The parameters $p$ and $k$ bear the signature of nonconformality. It is worth mentioning that the dilaton in this theory has a logarithmic profile given by $\phi=\phi_0-a_0\log r$. In the above metric, $\eta$ is the nonconformal parameter which determines the deviation of the dual field theory from the conformal fixed point. This nonconformal parameter obeys the bound $\eta^2<\frac{8d}{d-1}$ \cite{gubser}. At a special limit ($\eta\rightarrow 0$), the above black-brane solution reduces to the Schwarzschild black hole spacetime geometry and correspondingly the conformal symmetry is restored in the boundary field theory. It is therefore evident that this  solution of the Einstein-dilation theory makes it possible to holographically probe things in a more general set up.
\noindent The Hawking temperature of the above black brane geometry reads
\begin{eqnarray}\label{3}
T_{H} =  \left(\frac{8+(d-1)\eta^2}{8d+(d-1)\eta^2}\right)\frac{4\pi}{r_h^{\left(\frac{8-(d-1)\eta^2}{8+(d-1)\eta^2}\right)}}~.
\end{eqnarray}
In order to compute the HEE, we assume that the subsystem $A$ has a strip geometry specified as $-\frac{l}{2} <x_1 < \frac{l}{2}$ and $-\frac{L}{2}<x_{2;3;4;...;d-1}<\frac{L}{2}$. This in turn fixes the volume of the subsystem at the boundary field theory to be $V=L^{d-2}l$. The thermal entropy of the boundary field theory or more precisely the amount of the Bekenstein-Hawking entropy \cite{bek, BCH} captured in the above mentioned subsystem volume reads
$S_{BH}=\frac{L^{d-2}l}{4G_N}r_h^{p(d-1)}$. 
We now move on to compute the HEE by incorporating the Ryu-Takayanagi (RT) formula
$S_{E} = \frac{A(\gamma_A)}{4G_{N}}$ \cite{rt}.
By following the standard procedure of extremizing the area functional of the static minimal surface ($A(\gamma_A)$) and using the RT formula, we obtain the HEE of the Einstein-dilaton black brane to be
\begin{widetext}
	\begin{eqnarray}\label{se}
	S_E =  \frac{2L^{d-2}}{4G_N}\left[ \frac{\Lambda^{p(d-3)+1}}{[p(d-3)+1]}- \frac{r_t^{p(d-3)+1}}{[p(d-3)+1]}\left[\frac{\Gamma\left[\frac{p(d+1)-1}{2p(d-1)}\right]}{\Gamma\left[\frac{2p-1}{2p(d-1)}\right]}\right] +  \sum_{n=1}^{\infty} \frac{\left(\frac{r_h}{r_t}\right)^{nk}r_t^{p(d-3)+1}}{[2p(d-1)]}
	\left[\frac{\Gamma[n+\frac{1}{2}]\Gamma\left[\frac{nk-1-p(d-3)}{2p(d-1)}\right]}{\Gamma[n+1]\Gamma\left[\frac{2p-1+nk}{2p(d-1)}\right]}\right]\right]~~~
	\end{eqnarray}
\end{widetext}
\noindent where $\Lambda$ is a cut-off to prevent the universal divergence  of the entanglement entropy of the dual field theory and $r_t$ is the turning point. In the gauge/gravity set up, $S_E$ corresponds to the entanglement entropy of a thermally excited pure state in a $d$-dimensional field theory. On a similar note, the entanglement entropy of the ground state in a $d$-dimensional field theory can be holographically computed by using the asymptotic form ($r_h\rightarrow0$) of the black brane given in eq.(\ref{1}). 
We denote this quantity as $S_G$.
The expressions of $S_E$ (eq.(\ref{se})) and $S_G$ enable us to compute the change in the HEE due to thermal excitation of the dual field theory. We call this change as the renormalized HEE ($S_{REE} = S_E-S_G$) since it is a finite quantity. Now it is well known that  black holes and black branes are thermodynamical systems and the first law of black hole thermodynamics reads 
$dE = T_H~dS_{BH}~$ \cite{BCH}. This leads to the following change in the internal energy due to the formation of black brane in the Einstein-dilaton gravity\\
\begin{eqnarray}\label{14}
E = \int_{0}^{r_h} T_H~dS_{BH} = \left[\frac{8(d-1)}{8d-(d-1)\eta^2}\right] T_H~S_{BH}~.
\end{eqnarray}
This is the Smarr relation of black hole thermodynamics \cite{smarr}. In the subsequent discussion, we shall see that one can obtain a thermodynamics like law in the context of HEE using the above expression of internal energy ($E$) and the renormalized HEE ($S_{REE}$). We shall find that this relation is identical with the Smarr relation of black hole thermodynamics (eq.\ref{14}). To proceed, we introduce a generalized temperature $T_g$,  valid for all possible values of the subsystem size $l$, as
\begin{eqnarray}\label{10}
\frac{1}{T_g} =\lambda \frac{S_{REE}}{E}
\end{eqnarray}
where $\lambda$ is a proportionality constant which we shall fix by demanding that $T_g$ yields the Hawking temperature of the black hole in the IR limit. The justification for demanding this condition is the following. In the IR limit ($l\rightarrow\infty$), the static minimal surface wraps a portion of the event horizon of the black hole in the bulk, which suggests that in the $l\rightarrow\infty$ limit, $T_g$ must reduce to the Hawking temperature $T_H$ of the black brane. On the other hand, in the $l\rightarrow 0$ limit, $T_g$ would yield the entanglement temperature of the boundary field theory. In the large $l$ limit, eq.(\ref{10}) reads
\begin{eqnarray}\label{11}
\frac{1}{T_g}&=&\lambda\left[\frac{8d-(d-1)\eta^2}{8(d-1)}\right]\frac{1}{T_H}\nonumber\\
&&\times\left[1+\frac{\Xi_1}{lr_h^{\left(\frac{8-(d-1)\eta^2}{8+(d-1)\eta^2}\right)}}+\frac{\Xi_2}{r_h^{p(d-1)}l^{\frac{p(d-1)}{2p-1}}}+...\right]
\end{eqnarray}
\noindent where $\Xi_1$ and $\Xi_2$ are real numbers which depend upon the spacetime dimensions and the nonconformal parameter $\eta$.
Demanding that $T_g$ should reduce to the Hawking temperature in the $l\rightarrow\infty$ limit, fixes the value of the proportionality constant $\lambda$ to be $\lambda=\frac{8(d-1)}{8d-(d-1)\eta^2}$.
It is worth mentioning that the terms with coefficient $\Xi_1$ and $\Xi_2$ are smaller in magnitude in the large $l$ limit as they scale with the inverse of the subsystem size $l$. These terms are the correction terms and bear the signature of the short distance correlation along the entangling surface in the dual field theory. Fixing the value of the proportionality constant $\lambda$ in eq.(\ref{11}), we get
\begin{eqnarray}
\frac{1}{T_g} = \frac{1}{T_H} \left[1+\mathrm{correction~terms}\right]~.
\end{eqnarray}
The form of the thermodynamics like law for the HEE given in eq.(\ref{10}) therefore reads
\begin{eqnarray}\label{12}
E = \left[\frac{8(d-1)}{8d-(d-1)\eta^2}\right] T_g ~S_{REE}~.
\end{eqnarray}
Interestingly, it can be observed that the above relation matches exactly with the Smarr relation of black hole thermodynamics (eq.(\ref{14})). This is one of the main findings in this letter. 
Hence, this relation can be called as the Smarr relation for HEE.
We emphasize the fact that the Smarr relation satisfied 
by the HEE has been obtained from the physical requirement that the generalized temperature yields the Hawking temperature in the IR limit.
It is reassuring to note that for $\eta=0$, we recover the result obtained in \cite{prd}. Now we move on to the UV limit ($l\rightarrow 0$). In this limit, $T_g$ gives the entanglement temperature. This reads
   \begin{eqnarray}\label{13}
	\frac{1}{T_g} &=&
\Xi_3 l^{\frac{k-p(d-1)}{2p-1}}\times\left[1+\Xi_4\left(r_h~l^{\frac{1}{2p-1}}\right)^k +...\right]
	\end{eqnarray}
where $\Xi_3$ and $\Xi_4$ are real numbers which depend upon the spacetime dimensions and the nonconformal parameter $\eta$.
Eq.(\ref{13}) shows that the leading term of the generalized temperature $T_g$ yields the well-known inverse of the subsystem size in the UV domain. The next term is the sub-leading correction to the entanglement temperature. The Smarr relation for HEE also  probes the initiation of thermalization in the dual field theory as we move from UV domain to the IR domain. This has been graphically represented in \cite{prd}
where the flow of the generalized entanglement temperature $T_g$ with the subsystem size $l$ has been represented. The flow diagram reveals the initiation of the thermalization in the dual field theory at a certain value $l_c$ of the subsystem size.

We now present another example of a spacetime (called the hyperscaling violating geometry) dual to nonrelativistic critical points. The metric in this case reads \cite{hyper}
	\begin{eqnarray}
	ds^2= \frac{R^2}{r^2}\left[-\frac{f(r)}{ r^{\frac{2(d-1)(z-1)}{(d-\theta-1)}}}dt^2 + r^{\frac{2\theta}{d-\theta-1}}\frac{dr^2}{f(r)}+\sum_{i=1}^{d-1}dx_i^2\right].\nonumber\\
	\end{eqnarray} 
The lapse function is given as $f(r)= 1-\left(\frac{r}{r_h}\right)^{(d-1)(1+\frac{z}{d-\theta -1})}$, where $z$ is the dynamical exponent and $\theta$ is the hyperscaling violating exponent. This geometry represents holographic theories with hidden Fermi surfaces. The change in the internal energy in this case is obtained to be
\begin{eqnarray}
E = \left(\frac{d-\theta-1}{z+d-\theta-1}\right)T_H~S_{BH}
\label{hy}
\end{eqnarray}
where the black hole entropy $S_{BH}=\frac{L^{d-2} l}{4G_N r_{h}^{d-1}}$ \cite{hyper} and the Hawking temperature $T_H$ is given by
\begin{eqnarray}
T_H =\frac{(d-1)(z+ d-\theta -1)}{4\pi (d-\theta -1)}\frac{1}{r_{h}^{z(d-1)/(d-\theta -1)}}~.
\label{H}
\end{eqnarray}
Once again we introduce a generalized temperature $T_g$ 
in this case as in eq.(\ref{10}). Remembering that $T_g =T_H$ in the IR limit ($l\rightarrow\infty$) fixes the proportionality constant $\lambda$ in eq.(\ref{10}) to be $\lambda=\left(\frac{d-\theta-1}{z+d-\theta-1}\right)$. This immediately leads to the thermodynamics like law for HEE to be
\begin{eqnarray}\label{15}
E = \left(\frac{d-\theta-1}{z+d-\theta-1}\right)T_g~S_{REE}~.
\end{eqnarray}
This is exactly identical to the Smarr relation eq.(\ref{hy}).
In the UV limit, $T_g$ gives the entanglement temperature, which reads
\begin{eqnarray}
\frac{1}{T_g} = \Xi_5 ~l^z
\end{eqnarray}
where $\Xi_5$ is a number depending upon the spacetime dimensions and the parameters involved in the theory. The above result is consistent with previous finding \cite{hyper2}.

So far we have considered gravity backgrounds which are holographically dual to relativistic quantum field theories with vanishing chemical potential. A natural question which arises is whether a thermodynamics like law satisfied by the HEE exists for a dual gravity background with charge. We find that the answer to this question is in the affirmative. To show this we consider a AdS$_{d+1}$-Reissner-Nordstr\"om (RN) black hole with charge $q$ which is dual to a $d$-dimensional field theory with non-zero chemical potential. The metric of this background reads (with AdS radius $R=1$)
\begin{widetext}
\begin{eqnarray}\label{RN}
	ds^2 = \frac{1}{z^2}\left[-f(z)dt^2 + \frac{dz^2}{f(z)}+\sum_{i=1}^{d-1} dx_i^2\right];~
	f(z) = 1-\left(\frac{z}{z_h}\right)^d -  \frac{(d-2)}{(d-1)}q^2z_h^{2(d-1)} \left(\frac{z}{z_h}\right)^d \left(1-\left(\frac{z}{z_h}\right)^{d-2}\right). 
\end{eqnarray}
\end{widetext}
For static charged black holes, the first law of black hole thermodynamics reads $dM=T_H dS_{BH} + \phi dq$, where $\phi$ is the electrostatic potential and $q$ represents the total charge \cite{BCH}. 
For AdS-planar charged black holes, the above law leads to the following thermodynamic Smarr relation \cite{pope}
\begin{eqnarray}\label{ern}
M= \left(\frac{d-1}{d}\right)\left[T_HS_{BH}+\phi q\right]~.
\end{eqnarray}
It is  worth mentioning that the above formula is obtained by considering the system in a grand canonical ensemble and noting that pure AdS is the ground state of the dual field theory\footnote{Further details in this direction 
can be found in \cite{mayers}. One can also consider the system in a canonical ensemble, in that case the ground state of the dual field theory 
is holographically represented by the 
extremal limit of the AdS-RN black hole \cite{kundu}.}.
In \cite{cai}, it was argued that in order to obtain the internal energy $E$ of the AdS-RN black hole, the contribution of the static electricity part should be subtracted. Therefore, the internal energy $E$ of a $(d+1)$-dimensional planar AdS-RN black hole reads $E = M-\left(\frac{d-1}{d}\right) \phi q$. Hence, the change in the internal energy which is obtained from the difference in the internal energy of the charged black hole from the pure AdS$_{d+1}$ spacetime reads
\begin{eqnarray}\label{16}
E = \left(\frac{d-1}{d}\right)T_HS_{BH}
\end{eqnarray} 
where the black hole entropy $S_{BH}=\frac{L^{d-2}l}{4G_N z_{h}^{d-1}}$ and the Hawking temperature $T_H$ is given by
\begin{eqnarray}\label{16ab}
T_H = \frac{d}{4\pi z_h}\left[1-\frac{(d-2)^2 q^2 z_{h}^{2(d-1)}}{d(d-1)}\right].
\end{eqnarray} 
Similar to the previous examples, we introduce a generalized temperature $T_g$ as in  eq.(\ref{10}). The internal energy $E$ is given in eq.(\ref{16}), and the renormalized holographic entanglement entropy $S_{REE}$ in this case reads $S_{REE} = S_{non-ext} - S_{AdS_{d+1}}$, where $S_{non-ext}$ corresponds to the HEE of the non-extremal AdS-RN black hole. As we have mentioned earlier, we fix the proportionality constant $\lambda$ by demanding that $T_g$ should reduce to the Hawking temperature $T_H$ in the IR limit.
We now restrict to the small charge limit of the black hole ($\alpha \equiv \sqrt{\frac{d-2}{d-1}} q z_h^{d-1}\ll1$) since in the large charge limit ($\alpha\gg1$), the turning point $z_t$ approaches the event horizon $z_h$ irrespective of the thermal property of the system (for both extremal and non-extremal cases) \cite{kundu}. In the large $l$ limit, eq.(\ref{10}) yields 
\begin{eqnarray}
\frac{1}{T_g} = \lambda\left(\frac{d}{d-1}\right)\left[\frac{1}{T_H}+...\right].
\end{eqnarray}
The proportionality constant $\lambda$ in this case gets fixed to the value 
$\lambda = \left(\frac{d-1}{d}\right)$. This in turn leads to the following form of the thermodynamics like law for HEE
\begin{eqnarray}\label{17}
\frac{1}{T_g} = \left(\frac{d-1}{d}\right)\frac{S_{REE}}{E}
\end{eqnarray}
which is exactly identical to the Smarr relation (eq.(\ref{16})). 
In the UV limit ($l\rightarrow 0$), eq.(\ref{17}) leads to the entanglement temperature up to $\mathcal{O}(\alpha^2)$ to be
\begin{eqnarray}\label{18}
\frac{1}{T_g} = \Xi_6 l \left[1+ \frac{2(d-1)}{d}\alpha^2\right]
\end{eqnarray}
where $\Xi_6$ is a constant term which only depends upon the spacetime dimensions. It is interesting to observe that the entanglement temperature in this case gets a correction due to the presence of the charge parameter $\alpha$. Furthermore, the 
leading term in the right side of eq.(\ref{18}) is independent of $\alpha$ and depends only on the subsystem size $l$, and perfectly matches with that of SAdS$_{d+1}$ \cite{prd}.

We now summarize our findings. In this letter we obtain a Smarr like thermodynamic relation in the context of HEE. Remarkably, this relation is found to be exactly identical with the Smarr relation of black hole thermodynamics.  The relation arises by introducing the notion of a generalized temperature $T_g$ which is required to yield the Hawking temperature $T_H$ in the IR limit ($l\rightarrow\infty$). The way in which $T_g$ is introduced is quite different from the concept of the entanglement temperature $T_{ent}$, defined and valid only in the UV limit \cite{jb}. We demonstrate this for three different examples, namely, a relativistic gravity background with a nonconformal parameter, the hyperscaling violating geometry, and the charged AdS$_{d+1}$
background geometry dual to a field theory having a finite chemical potential. The charged black hole geometry is a bit more tricky as one needs to properly identify the internal energy by subtracting the static electricity part from the mass of the black hole. The Smarr relation for the HEE also yields the entanglement temperature proportional to the inverse of the subsystem size $l$ in the UV limit ($l\rightarrow 0$). Furthermore, we emphasize that the numerical factor of the leading term is correctly determined since it has been obtained from the generalized temperature $T_g$ which has been calibrated with the Hawking temperature of the black hole in the IR limit.
Finally, we would like to mention that deriving the Smarr like relation for entanglement thermodynamics from the Iyer-Wald formalism \cite{wald} would be very interesting. However, to obtain this one would need the entanglement first law (in differential form) which still remains an open problem to find for black holes \cite{cac}.

\end{document}